\documentclass[aps,
twocolumn,
showpacs,preprintnumbers,amsmath,amssymb,superscriptaddress,prx,floatfix,10pt]{revtex4-2}
\usepackage{graphicx}
\usepackage{float}
\usepackage{mathtools}
\usepackage{bbm}
\usepackage[normalem]{ulem}
\usepackage{xcolor}
\usepackage{comment}
\usepackage{amsthm} 
\usepackage{amssymb}   
\usepackage{cancel}
\usepackage[colorlinks, linkcolor=blue, citecolor=blue]{hyperref}

\begin{document}
\title{Towards a microscopic model for an electronic quantum charge liquid}

\author{Jacob R. Taylor}
\affiliation{Condensed Matter Theory Center and Joint Quantum Institute, Department of Physics, University of Maryland, College Park, Maryland 20742, USA}
\author{Sankar Das Sarma}
\affiliation{Condensed Matter Theory Center and Joint Quantum Institute, Department of Physics, University of Maryland, College Park, Maryland 20742, USA}
\author{Seth Musser}
\email[]{swmusser@gmail.com}
\affiliation{Condensed Matter Theory Center and Joint Quantum Institute, Department of Physics, University of Maryland, College Park, Maryland 20742, USA}

\begin{abstract}
We provide a route to constructing an electronic quantum charge liquid (QCL), a state made up of fermions at fractional filling of a lattice that does not break translation. Starting with spinless fermions at filling $\nu=3/2$ we pair them to get bosons at filling $\nu=3/4$ per unit cell. The tetramer model, a generalization of the dimer model, on the square lattice is evaluated as a candidate bosonic QCL at filling $\nu = 3/4$. It is shown that these models exhibit a local $\mathbb{Z}_4$ symmetry. Upon numerical study of a family of tetramer wavefunctions it is found that while one is gapless due to $\mathrm{U}(1)^3$ symmetry at least one other can be definitively shown to be gapped. The gapped nature of this state, along with its $\mathbb{Z}_4$ symmetry, leads us to propose that it is an example of the elusive bosonic QCL displaying the minimal $\mathbb{Z}_4$ topological order. We conclude by discussing possible extensions to other lattice geometries, electronic QCLs, and to Rydberg atoms.
\end{abstract}

\maketitle

\begin{figure}
    \centering
    \includegraphics[width=0.95\columnwidth]{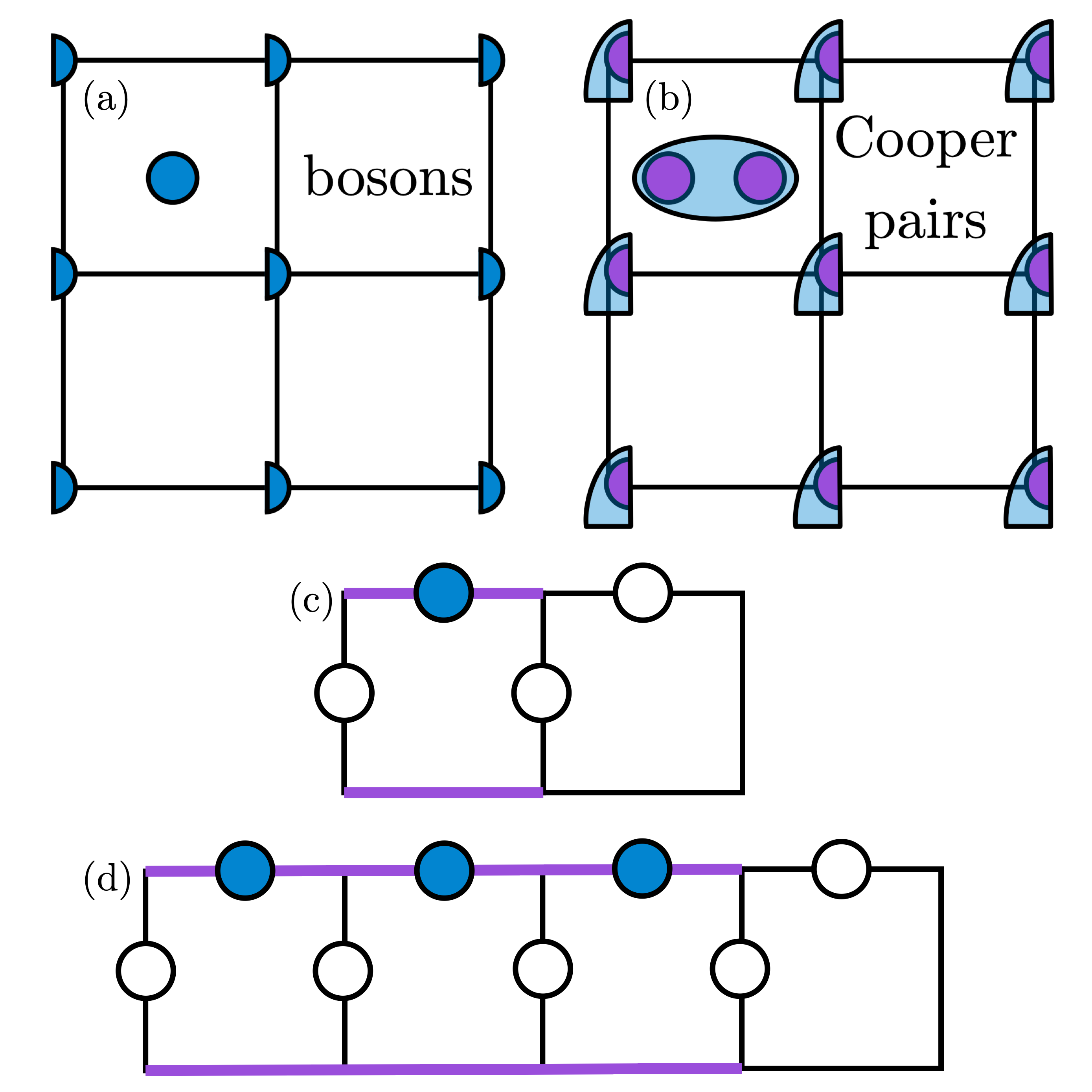}
    \caption{A visual explanation of minimal quantum charge liquids. (a) A system of charge-$1$ bosons at $\nu=1/2$ filling per unit cell that does not break translation requires a half-charged background anyon per unit cell \cite{cheng_translational_2016}. This is illustrated with the half-circle background anyons, which square to the circular elementary boson. (b) For a system of charge-$1$ spinless fermions at $\nu = 1/2$ a charge-$2$ boson can be made by Cooper pairing. To get a half-charged background anyon then requires taking a quarter of the Cooper pair. Four background anyons are required to return to the bosonic vacuum, leading to the $\mathbb{Z}_4$ toric code in this case. (c) An illustration of why the dimer model can be viewed as a hopping model of bosons at $\nu = 1/2$. The hardcore bosons, in blue, live on the bonds of the square lattice. An occupied site corresponds to a dimer on that bond, in purple. The extended unit cell of the columnar phase is shown, with an occupation of $\nu=1/2$ per unit cell. (d) An illustration of why the tetramer model can be viewed as a hopping model of bosons at $\nu=3/4$. Each link of the tetramer corresponds to an occupied site. In the columnar phase, pictured, this leads to a filling of $\nu=3/4$.}
    \label{fig:Z4_reason}
\end{figure}

\textbf{Introduction:} Strongly correlated electronic matter often relies on a balance between interelectron Coulomb repulsion and kinetic energy. When Coulomb repulsion dominates electrons prefer to form crystalline insulating states. If the electrons live on a crystalline lattice at some fractional filling per unit cell, $\nu$, then these insulating states are expected to break the lattice translation symmetry and are referred to as Wigner-Mott insulators. As an example, recent experiments \cite{regan_mott_2020, wang_correlated_2020, xu_correlated_2020, mak_semiconductor_2022} on moir\'{e} transition metal dichalcogenide (TMD) materials have found evidence for Wigner-Mott insulating phases at many fractional fillings $\nu$. In contrast, if kinetic energy dominates then metallic Fermi-liquid phases are expected. If the competition between these two energy scales is tuned, e.g. by a displacement field in the TMD materials, then it is possible to tune between the Wigner-Mott (or Mott if $\nu=1$) and metallic phases \cite{PanWuDasSarma2020, PanDasSarma2021, tang_dielectric_2022, kim_continuous_2023}.
A few works have treated the possibility of direct continuous phase transitions between the Wigner-Mott and metallic phases~\cite{xu_interaction-driven_2022, musser_theory_2022, musser_metal_2022}. However, it may be possible that another phase intervenes in between the two.

Recently one of the authors proposed a ``quantum charge liquid'' (QCL) phase that may be intermediate between a charge-ordered Wigner-Mott insulator and a translation symmetric metal. This QCL phase is presumed to inherit the translation symmetry of the metallic phase, while developing a charge gap \cite{musser_fractionalization_2025, krishnan_defect_2025}. We can then use the Lieb-Schultz-Mattis-Oshikawa-Hastings (LSMOH) theorem \cite{lieb_two_1961, oshikawa_commensurability_2000, hastings_lieb-schultz-mattis_2004} to see that this exotic QCL will necessarily have fractional-charge excitations and associated topological order (TO), or have gapless neutral excitations.

In fact, more can be proven about the QCL than the mere existence of TO. 
In Ref.~\cite{musser_fractionalization_2025} it was shown that if a bosonic QCL: 1) displays a gap to all excitations, 2) does not break time reversal, and 3) is at filling $\nu = p/q$ of a lattice, with $p$ and $q$ coprime, then it must have at least $q^2$ anyons. The unique TO achieving this lower bound, or minimal TO, is the $\mathbb{Z}_q$ toric code. By contrast in a fermionic QCL with the same restrictions the minimal number of anyons is $q^2$ ($4q^2$) for $q$ odd (even) and the unique minimal TO is the $\mathbb{Z}_q$ ($\mathbb{Z}_{2q}$) toric code, respectively. The proof of these statements relies on the existence of an Abelian background anyon, established in Ref.~\cite{cheng_translational_2016}. A visual explanation of the proof, along with why the fermionic case is different is illustrated in Fig.~\ref{fig:Z4_reason}(a-b). These generic formal results provide a background for our specific work in the current paper.

There are microscopic models of bosonic QCLs which, to our knowledge, all realize the minimal TO (if they are gapped). One early example is the square lattice quantum dimer model \cite{RokhsarKivelson1988}. If the dimers are mapped to hardcore bosons on the  bonds of the square lattice then the dimer model can be viewed as a hopping model of bosons at filling $\nu=1/2$. See Fig.~\ref{fig:Z4_reason}(c) for an illustration. Moreover, at the resonating valence bond (RVB) point the wavefunction does not break translation. By the LSMOH theorem it must have TO or be gapless, and the square lattice dimer model is indeed gapless \cite{RokhsarKivelson1988, ArdonneFendleyFradkin2004}. On the other hand the triangular lattice quantum dimer model, which can be mapped to a hopping problem of hardcore bosons on a kagome lattice \cite{moessner_quantum_2008}, is gapped and realizes the minimal $\mathbb{Z}_2$ TO \cite{MoessnerSondhi2001}. Other tractable microscopic models for bosonic fractionalized phases have also been constructed \cite{balents_fractionalization_2002,senthil_microscopic_2002, motrunich_exotic_2002}, mostly also realizing the minimal $\mathbb{Z}_2$ TO. In contrast fillings $\nu = p/q$ with $q>2$ have received less attention since they do not admit a description in terms of a simple parton mean field \cite{ye_classification_2024}. A handful of works have studied the case of $q=3$ using a variety of miscoscopic models \cite{motrunich_bosonic_2003, myers_z_3_2017, lee_resonating_2017, devakul_z_2018, dong_su3_2018, kurecic_gapped_2019, jandura_quantum_2020, giudice_trimer_2022, mao_fractionalization_2023, chen_how_2025}, with all topologically ordered phases found being the minimal $\mathbb{Z}_3$ TO. Many of these involved using a generalization of the dimer model, the trimer model. To our knowledge there have not been any microscopic models of bosonic QCLs with $q>3$ studied.

Microscopic models of electronic QCLs are even less explored than bosonic QCLs. Examples at filling $\nu=1$ involve using phonons to pair nearby electrons into Cooper pairs, leading to a bosonic dimer model that realizes $\mathbb{Z}_2$ TO \cite{han_resonating_2023, cai_quantum_2024}. These examples take advantage of Cooper pairing, illustrated in Fig.~\ref{fig:Z4_reason}(b), to reduce the electronic QCL to a bosonic QCL at a smaller filling. We are not aware of efforts to produce microscopic models of electronic QCLs at fractional fillings.

In this paper we address this question for the smallest denominator fractional filling, $q=2$, and choose $\nu = 3/2$. If we Cooper pair the electrons, as in the references above, then this reduces to studying bosonic QCLs at filling $\nu=3/4$. By analogy with the dimer and trimer models we consider a tetramer model on the square lattice. As Fig.~\ref{fig:Z4_reason}(d) shows, a na\"{i}ve mapping of tetramers to hardcore bosons will indeed lead to a hopping model of bosons at filling $\nu = 3/4$ per unit cell. Motivated by this we study a family of modified RVB wavefunctions for tetramers. We show that these wavefunctions always display a local $\mathbb{Z}_4$ flux conservation, and at a specific point this symmetry is enlarged to $\mathrm{U}(1)^3$. Mapping the wavefunction to a transfer matrix it is possible to numerically analyze the gap/gaplessness of these wavefunctions. We find that the $\mathrm{U}(1)^3$ symmetric point is gapless, as it should be \cite{polyakov_quark_1977}. However, tuning away from this point we find a gapped wavefunction with exponentially decaying correlation functions. This, in conjunction with the $\mathbb{Z}_4$ flux conservation, provides strong evidence that the modified tetramer model realizes the minimal $\mathbb{Z}_4$ TO. We conclude by discussing extensions of this work to other lattice geometries, more concrete microscopic models for electronic QCLs at $\nu = 3/2$, and to bosonic Rydberg atoms at filling $\nu = p/4$ with $p$ odd.

\begin{figure}
    \centering
    \includegraphics[width=\columnwidth]{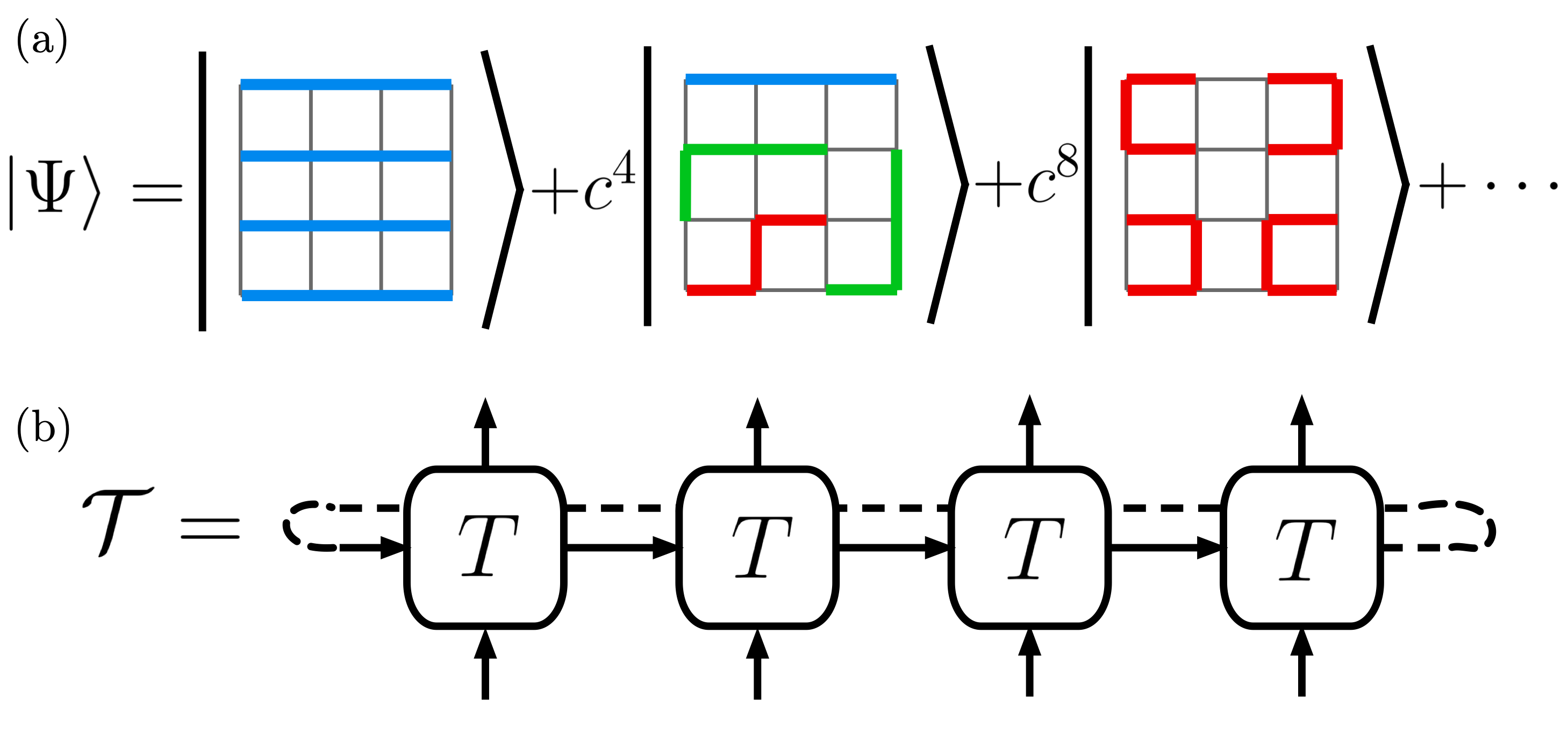}
    \caption{The wavefunction and transfer matrix. (a) The RVB wavefunction for the tetramer model on a $4\times 4$ lattice. Straight tetramers are colored blue, partially bent tetramers are colored green, and fully bent tetramers are colored red. Each tetramer configuration is weighted by a factor of $c^n=\cot^n \theta$, where $n$ is the total number of bends in the configuration (with blue tetramers having none, green having one, and red having two bends). (b) The transfer matrix, ${\cal T}$, for a semi-infinite cylinder with a circumference of four sites. The transfer matrix is built out of the rank-$4$ tensor $T$ applied on each site. The exact form of $T$ is given in Fig.~\ref{fig:tensor_network}. The dotted line indicates the periodic boundary conditions imposed.}
    \label{fig:wavefunction}
\end{figure}

\textbf{The model:} The one-parameter wavefunction we consider is illustrated in Fig.~\ref{fig:wavefunction}(a) for the $4\times 4$ square lattice. We take a weighted superposition of all tetramer configurations, inspired by a similar construction for the trimer model in Ref.~\cite{giudice_trimer_2022}. Up to rotation all possible tetramer shapes we consider are illustrated in Fig.~\ref{fig:wavefunction}(a). The weighting of each tetramer configuration is $c^n = \cot^n \theta$, where $n$ is the total number of bends in that configuration. Thus when $\theta=0$ only fully bent configurations, i.e. those with only red tetramers, are allowed, while when $\theta = \pi/2$ only fully straight configurations are, i.e. those with only blue tetramers. Explicitly, if $|t\rangle$ is a tetramer configuration then
\begin{equation}
|\Psi\rangle = \frac{1}{N(\theta)}\sum_{\{t\}} W(\theta,t)|t\rangle,
\end{equation}
where
\begin{equation}
W(\theta,t) = \left(\cos\theta\right)^{2N_{\mathrm{red},t}+N_{\mathrm{green},t}}\left(\sin\theta\right)^{2N_{\mathrm{blue},t}+N_{\mathrm{green},t}},
\end{equation}
is the weight and $N(\theta)$ is a normalization. Here $N_{\mathrm{red},t}$ is the number of fully bent, or red, tetramers in the configuration $|t\rangle$, likewise with the other colors.

This wavefunction can be represented as a projected entangled pair state (PEPS) on the semi-infinite cylinder using very similar constructions as those in Refs.~\cite{jandura2019topological, giudice_trimer_2022}. Once in PEPS, the correlation length of any operator is controlled by the eigenvalues of the transfer matrix, ${\cal T}$, one ring of the PEPS contracted with itself \cite{cirac_entanglement_2011, schuch_topological_2013, jandura2019topological}. See Fig. 1(e) of Ref.~\cite{schuch_topological_2013} for an illustration of this. For general $N$-mer models it is possible to directly determine the transfer matrix without reference to the PEPS. This was worked out in \cite{jandura2019topological} for generic values of $N$ and generic lattices. To build ${\cal T}$ for a generic lattice involves assigning a direction to each bond and placing tensors $T$ on each site whose rank is given by the coordination number of that site. For example on the square lattice $T$ is rank-$4$, written as $T_{ij\alpha\beta}$ and shown in Fig.~\ref{fig:wavefunction}(b). Here $i$ is the bottom leg, $j$ is the top, $\alpha$ the left, and $\beta$ the right with $i,j,\alpha,\beta \in \{0,1,2,3\}$.

\begin{figure}
    \centering
    \includegraphics[width=\columnwidth]{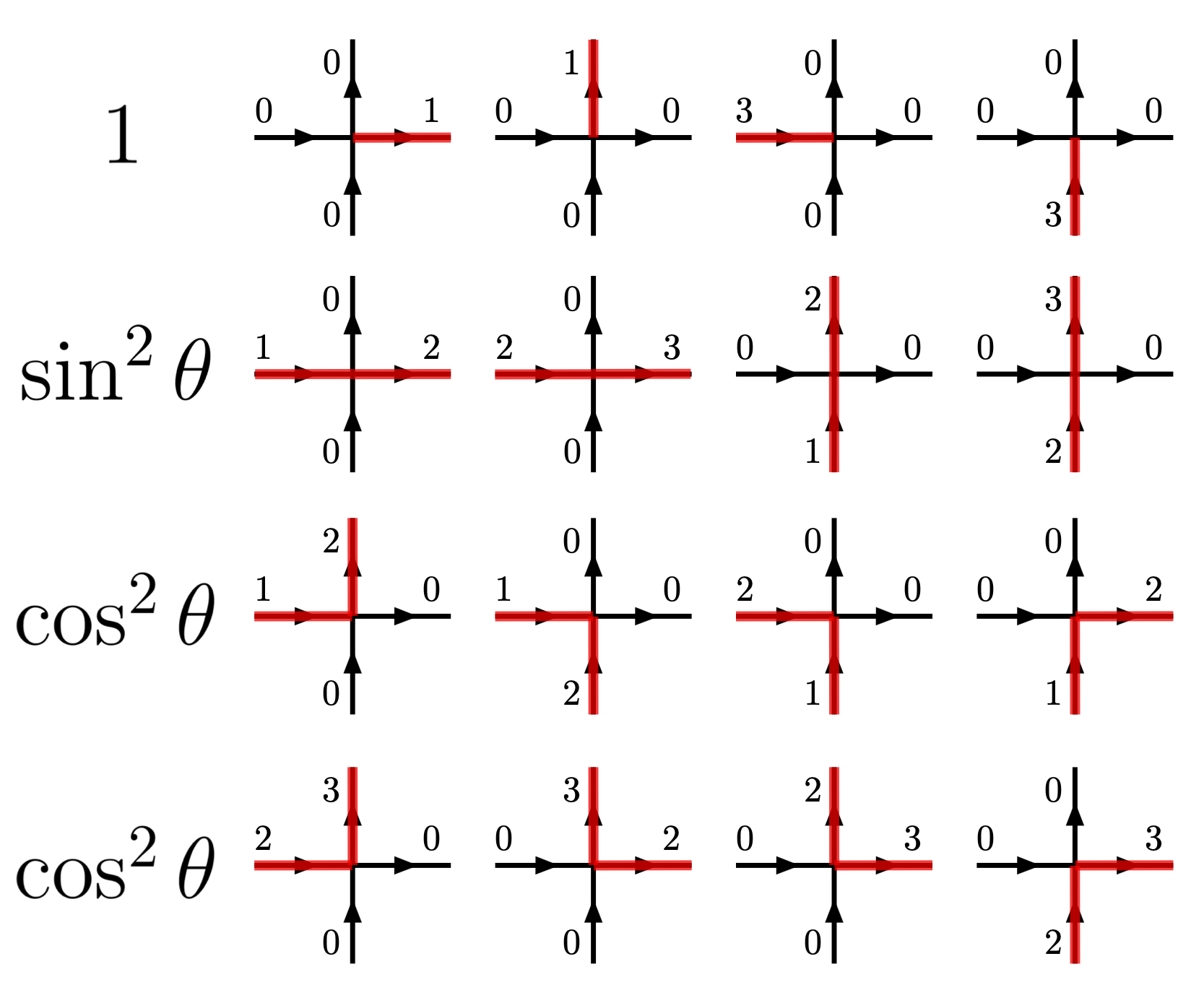}
    \caption{An explicit form of the tensor $T_{ij\alpha\beta}$ displayed in Fig.~\ref{fig:wavefunction}(b) and discussed in the main text. When $i,j,\alpha,\beta$ are one of the configurations displayed above then $T_{ij\alpha\beta}$ takes the value at left for that row. For all other $i,j,\alpha,\beta$, $T_{ij\alpha\beta}=0$. To find the flux $\pmod{4}$ emanating from a vertex subtract the value on an incoming arrow from the value on an outgoing arrow and take  the result $\pmod{4}$. All vertices have an outgoing flux of $1\pmod{4}$.}
    \label{fig:tensor_network}
\end{figure}

The specific form taken by $T$ for the wavefunction in Fig.~\ref{fig:wavefunction}(a) is shown in Fig.~\ref{fig:tensor_network}. There is a straightforward way to understand why $T$ allows only tetramer configurations by assigning values of $i,j,\alpha,\beta$ on bonds to tetramer links. If any of these values on the bonds is zero, then leave that bond blank, indicating no tetramer. However, if a bond value is nonzero then color the bond red, indicating a tetramer. From Fig.~\ref{fig:tensor_network} we see immediately that every site must be touched by one, and only one, tetramer. Further inspection shows that when we contract four $T$s on different sites, by pairing equivalent numbers together, only tetramer configurations will survive. The weighting on the left of Fig.~\ref{fig:tensor_network} assigns a weight of $\sin\theta$ to straight pieces of the tetramer and a weight of $\cos\theta$ to the bent pieces. These appear with a square because the transfer matrix can be thought of as a piece of the wavefunction contracted with itself, naturally leading to squared weights \cite{giudice_trimer_2022}.

\begin{figure*}
    \centering
    \includegraphics[width=0.9\textwidth]{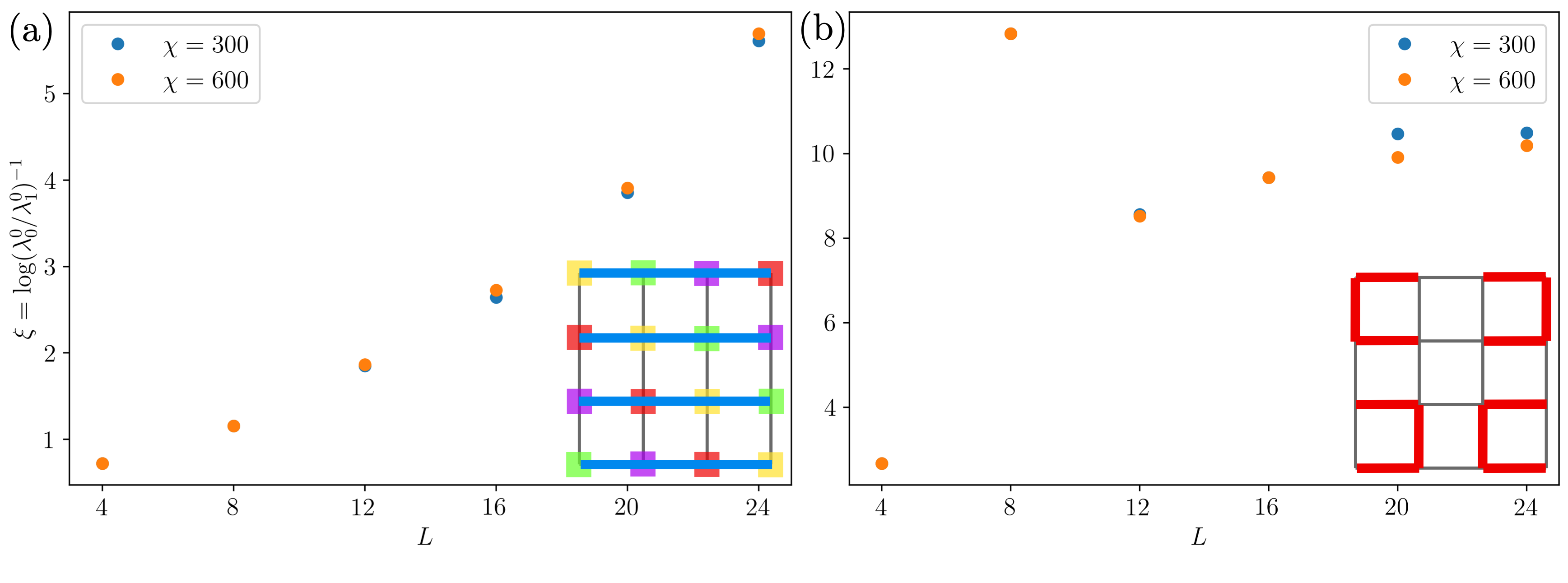}
    \caption{The upper bound on correlation length, $\xi$, vs. cylinder circumference, $L$, for fully straight and fully bent tetramer configurations, illustrated in the insets. (a) The upper bound on correlation length for fully straight configurations. As $L\rightarrow \infty$ we see that $\xi$ is diverging. Increasing the bond dimension from $\chi = 300$ to $\chi = 600$ enhances the divergence. These are indicative of gaplessness, coming from a $\mathrm{U}(1)^3$ symmetry. The symmetry can be seen by partitioning the lattice into four sublattices, indicated by the different colors in the inset, and is discussed in the main text. (b) The upper bound on correlation length for fully bent configurations. Now as $L\rightarrow \infty$ the correlation length $\xi$ appears to saturate. Increasing the bond dimension lowers $\xi$ at large circumference. These results indicate a gapped state.}
    \label{fig:correlation}
\end{figure*}

There is a $\mathbb{Z}_4$ local symmetry of the tetramer model that the transfer matrix makes clear. Every tetramer shape and orientation can be uniquely broken down into four of the on-site pieces shown in Fig.~\ref{fig:tensor_network}. We can therefore assign numbers taking values in $\{1,2,3\}$ to each of the links of every tetramer (and a value of $0$ to every unfilled bond). Further, these numbers obey a local $\mathbb{Z}_4$ flux conservation. This can be seen by inspection of the $16$ vertices in Fig.~\ref{fig:tensor_network}, where there is always a net outgoing flux of $1\pmod{4}$. Thus if we draw any path on the dual lattice encircling $N_v$ vertices, then we will enclose $N_v\pmod{4}$ flux. In the dimer (trimer) model local $\mathbb{Z}_2$ ($\mathbb{Z}_3$) flux conservation leads to $4$ ($9$) different equivalent sectors on a torus which are distinguished by their winding number \cite{moessner_quantum_2008, lee_resonating_2017, giudice_trimer_2022}. In an exactly analogous way this $\mathbb{Z}_4$ flux conservation will lead to (at least) $16$ different sectors on a torus. However, because we consider a semi-infinite cylinder we expect only a fourfold degeneracy in the spectrum of ${\cal T}$. Indeed we find numerically that for all values of $\theta$ the logarithm of the largest four eigenvalues of ${\cal T}$ have gaps that close exponentially in $L$, as expected for a transfer matrix with $\mathbb{Z}_4$ symmetry \cite{schuch_topological_2013}. We give more detail below.

We note that this local symmetry does not \textit{guarantee} topological order. For example, in the dimer model on the square lattice the local $\mathbb{Z}_2$ flux conservation is enlarged to a $\mathrm{U}(1)$ symmetry, leading to gaplessness and destroying the topological order \cite{moessner_quantum_2008}. Thus the remaining question is whether any of the wavefunctions we consider in Fig.~\ref{fig:wavefunction}(a) can be gapped; if they are then the observation of the $\mathbb{Z}_4$ symmetry will allow us to conclude the wavefunction realizes the minimal $\mathbb{Z}_4$ TO.

For the trimer model the question of gaplessness was addressed by Ref.~\cite{lee_resonating_2017} by measuring a variety of correlators at different length scales and concluding that all displayed exponentially decaying behavior, with a correlation length on the order of one to two lattice spacings. However, this left open the possibility that other unmeasured correlators may have power law behavior, indicating gaplessness. This was resolved by Ref.~\cite{giudice_trimer_2022} using the fact that at large distances \textit{all} correlation lengths are upper bounded by 
\begin{equation}
\xi = 1/\log \left|\frac{\lambda_0}{\lambda_1}\right|,
\end{equation}
where $\lambda_0$ is the largest eigenvalue of ${\cal T}$, $\lambda_1$ is the next largest, etc. We do the same for the tetramer model. As mentioned earlier we find a fourfold near degeneracy due to the $\mathbb{Z}_4$ flux conservation law, so the appropriate ratio to consider is $\lambda_0^0/\lambda_1^0$, where the superscript $0$ indicates that this is the charge $0$ sector of the $\mathbb{Z}_4$ symmetry. With this caveat in mind we find these eigenvalues numerically using tensor networks. In particular, we treat ${\cal T}$ as a matrix product operator (MPO) and iteratively extract its largest-magnitude eigenvalues and eigenstates using 10 iterations of a DMRG-like Arnoldi method \cite{arnoldi1951}. After obtaining an eigenpair $(\lambda_0,|\psi_0\rangle)$, we update the MPO to ${\cal T}'={\cal T}-\lambda_0 |\psi_0\rangle\langle\psi_0|$ (more precisely, only the action of $\lambda_0 |\psi_0\rangle\langle\psi_0|$ is used), and repeat the procedure to obtain subsequent eigenvalues. In practice, we compute 10 eigenvalues since Arnoldi often returns them slightly out of order. We note that despite the larger bond dimensions of the tetramer model compared to the trimer model we are still able to achieve larger system sizes than considered in \cite{lee_resonating_2017, giudice_trimer_2022}.
 
Our results are displayed in Fig.~\ref{fig:correlation}. We find that the fully straight tetramer wavefunction displays a diverging correlation length as $L\rightarrow \infty$, indicating gaplessness. The origin of gaplessness can be understood by partitioning the lattice into four sublattices, colored in the inset. Regardless of the orientation a straight trimer can be seen to contain one, and only one, of each color. We can then define three different ``electric field'' lines that originate from the yellow sublattice and end on the: green, purple, and red sublattices, respectively. The flux of the first field through any region will be given by the number of green sublattice points enclosed minus the number of yellow; likewise for the others. This leads to a flux valued in $\mathbb{Z}$, giving a local $\mathrm{U}(1)$ conservation law for each electric field. These conservation laws are independent, as can be seen by arguments nearly identical to those in App.~B of Ref.~\cite{giudice_trimer_2022}. Thus this state possesses a $\mathrm{U}(1)^3$ conservation law, leading to gaplessness \cite{polyakov_quark_1977}.

In contrast the fully bent tetramer wavefunction displays a saturating correlation length as $L\rightarrow \infty$. Increasing the bond dimension makes this saturating correlation length clearer. This implies a gapped state which, combined with the $\mathbb{Z}_4$ degeneracy of ${\cal T}$, suggests that the fully bent tetramer wavefunction exhibits the $\mathbb{Z}_4$ minimal TO for a bosonic state at this filling.

We note that we also measured $\xi$ versus $L$ for the wavefunction at $\theta= \pi/4$, i.e. the equal mixture state of all tetramers. There we find an apparently diverging correlation length. However, unlike $\theta=\pi/2$ there is no obvious conservation law leading to gaplessness. This leads us to speculate that this state is gapped, albeit with a very small gap. Thus the saturation of the correlation is not evident at the system sizes we have access to. As $\theta \rightarrow 0$, near the fully bent configuration, the gap becomes clear at the system sizes we can access. 

\textbf{Further work:} There are a number of possible extensions of this work. First it may be interesting to determine the correlation length in the infinite system limit by the use of tensor network methods such as CTMRG or VUMPS, as in \cite{giudice_trimer_2022}. Additionally we could consider the tetramer model on other lattice geometries. We consider the equal weighted tetramer wavefunction on the triangular lattice in App.~\ref{app:triang}, though we lacked the numerical resources to fully explore that problem. Other lattices, e.g. the kagome lattice, could also be studied.

Studying other lattice geometries may also be helpful to develop more concrete microscopic models for electronic QCLs. In the picture illustrated in Fig.~\ref{fig:Z4_reason}(d) each link of the tetramers we consider corresponds to a Cooper pair. The tetramer constraint is very unnatural when viewed from this lens, but may be more natural for tetramers in another lattice geometry. For example, in the trimer case Ref.~\cite{giudice_trimer_2022} showed that the all bent trimer wavefunction on the square lattice could be mapped to a boson hopping model on a two layer system. The trimer constraint was then naturally realized as a Rydberg blockade on this lattice. It would be interesting to see if other variants of the tetramer model on other lattice geometries can also have a natural physical realization while maintaining their $\mathbb{Z}_4$ minimal TO.

\acknowledgements
S.M. acknowledges helpful discussions with Andr\'{e} Grossi Fonseca, Jung Hoon Han, Hosho Katsura, Kyung-Su Kim, T. Senthil, and Victoria Zhang. This work is supported by the Laboratory for Physical Sciences through the Condensed Matter Theory Center.

\bibliographystyle{apsrev4-1_custom}
\bibliography{tetramer.bib}

\ \\ \pagebreak
\ \\ \pagebreak
\appendix

\section{Tetramers on the triangular lattice}\label{app:triang}

As in the main text we can map the tetramer wavefunction on the triangular lattice to a transfer matrix. We follow Eq.~(3) of \cite{jandura2019topological} to directly construct the transfer matrix. This is now a rank-$6$ tensor, $T_{ijkl\alpha\beta}$, since the triangular lattice has coordination number six. As in Fig.~\ref{fig:tensor_network} for the square lattice only some configurations of $i,j,k,l,\alpha,\beta$ give nonzero values of $T$. In total there are now $36$ configurations that give a nonzero value, so we do not plot them all here, but refer interested readers to Eq.~(3) of \cite{jandura2019topological}. We considered an equal weight superposition of all tetramer configurations, so we took any nonzero value of $T$ to be one rather than weighting by a factor related to the bending of a tetramer.

\begin{figure}
    \centering
    \includegraphics[width=\columnwidth]{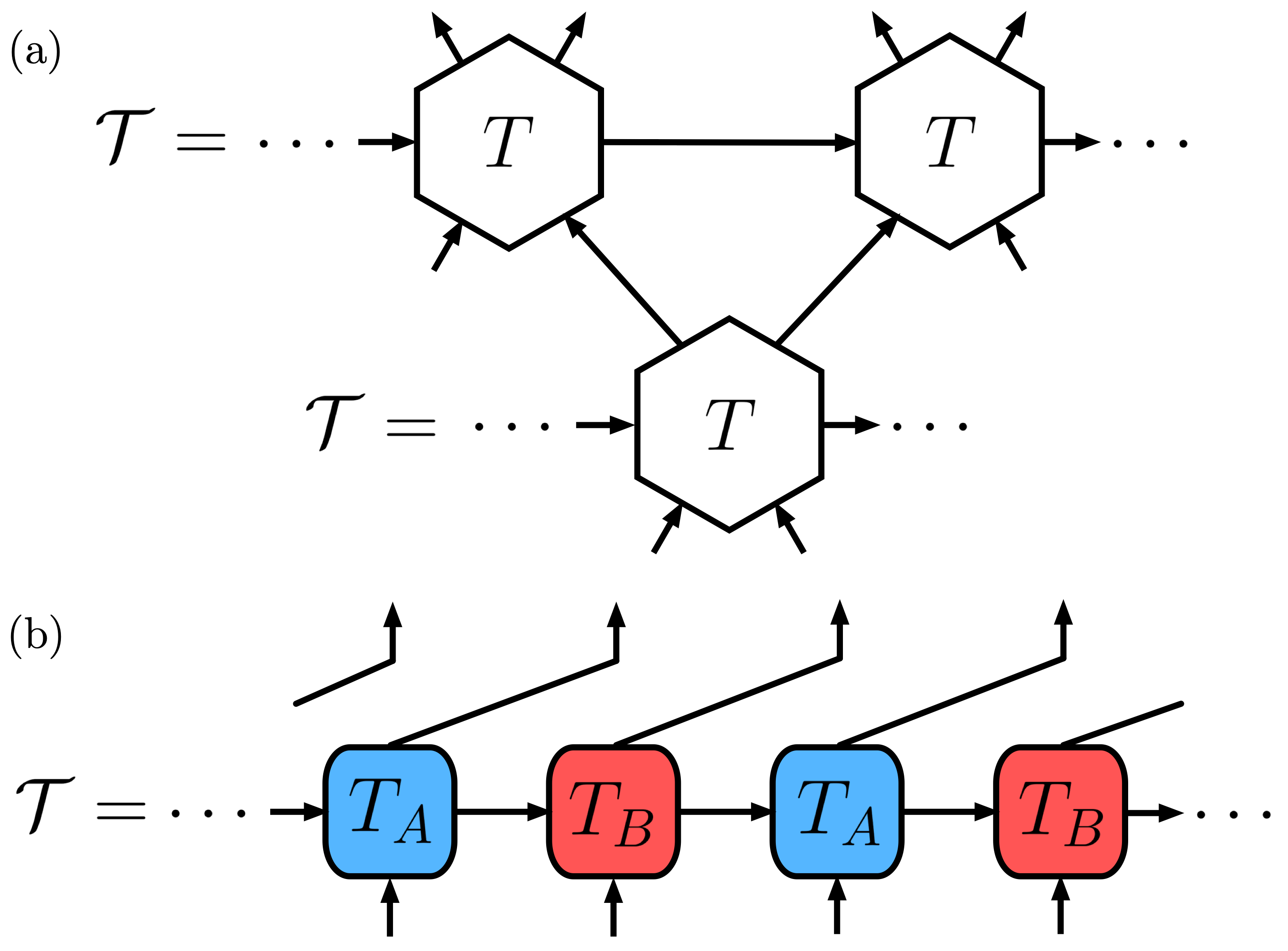}
    \caption{The transfer matrix and our numerical process for the triangular lattice. (a) The transfer matrix is built out of rank-$6$ tensors $T$. We show only a section of this. Because of the geometry of the triangular lattice different layers of the transfer matrix must be shifted by half a lattice site relative to one another. (b) For ease of computation the rank-$6$ tensor $T$ can be decomposed into two rank-$4$ tensors contracted together, $T_A$ and $T_B$. Thus by doubling the number of tensors we can return our transfer matrix to the square lattice form. In addition to this decomposition we must then add a shift to ensure $T_A$ is coupled to $T_B$ in the next layer.}
    \label{fig:trans_triang}
\end{figure}

The rank-$6$ tensors $T$ can be contracted along the legs labeled by Greek characters to form the transfer matrix, ${\cal T}$. We have plotted a segment of this in Fig.~\ref{fig:trans_triang}(a). Importantly the triangular case requires a shift by half a lattice site when stacking ${\cal T}$. This proves challenging to diagonalize, so instead we decompose $T$ into two rank-$4$ tensors $T_A$ and $T_B$ using a SVD decomposition. We are thus able to return our transfer matrix to the square lattice form, with this additional lattice shift. However, this transformation reveals the extra costs of diagonalizing ${\cal T}$ on the triangular lattice. It effectively doubles the number of lattice sites for a given cylinder circumference $L$, and also increases the bond dimension on the virtual legs.

Despite both of these challenges, the difficulty arises primarily because the MPO is not composed of a single $\mathcal{T}$, but of two layers. These MPOs cannot be freely stacked because of the geometry of the triangular lattice. Instead, for the first layer, the first site must be moved to the end through repeated swap operations, while for the second layer the reverse must be done, in order to construct a total MPO with periodic boundaries on the sites that can then be stacked. These repeated long-range swap operations substantially increase the bond dimension of the total MPO and scale very poorly with system size. For example, the MPO for a $4\times 4$ lattice had a bond dimension of $M=256$. Combined with the fact that the effective chain length is doubled, this makes the $4\times 4$ case comparable to solving a 40-site four-level chain with a very large-MPO bond dimension. In addition, when iteratively targeting higher excited eigenvalues, one must include the outer product of each previously obtained eigenstate, which further increases the computational cost. Because the MPS itself already has a moderate bond dimension, this process further enlarges the effective MPO at each eigenvalue iteration. In practice, the full MPS-MPO composite MPO is not constructed explicitly, but its action still carries a substantial computational cost. At an MPS bond dimension of 400, computing the largest 10 eigenstates required roughly 20 days for a cylinder circumference of $L=16$, making larger system sizes prohibitively expensive.

\begin{figure}
    \centering
    \includegraphics[width=\columnwidth]{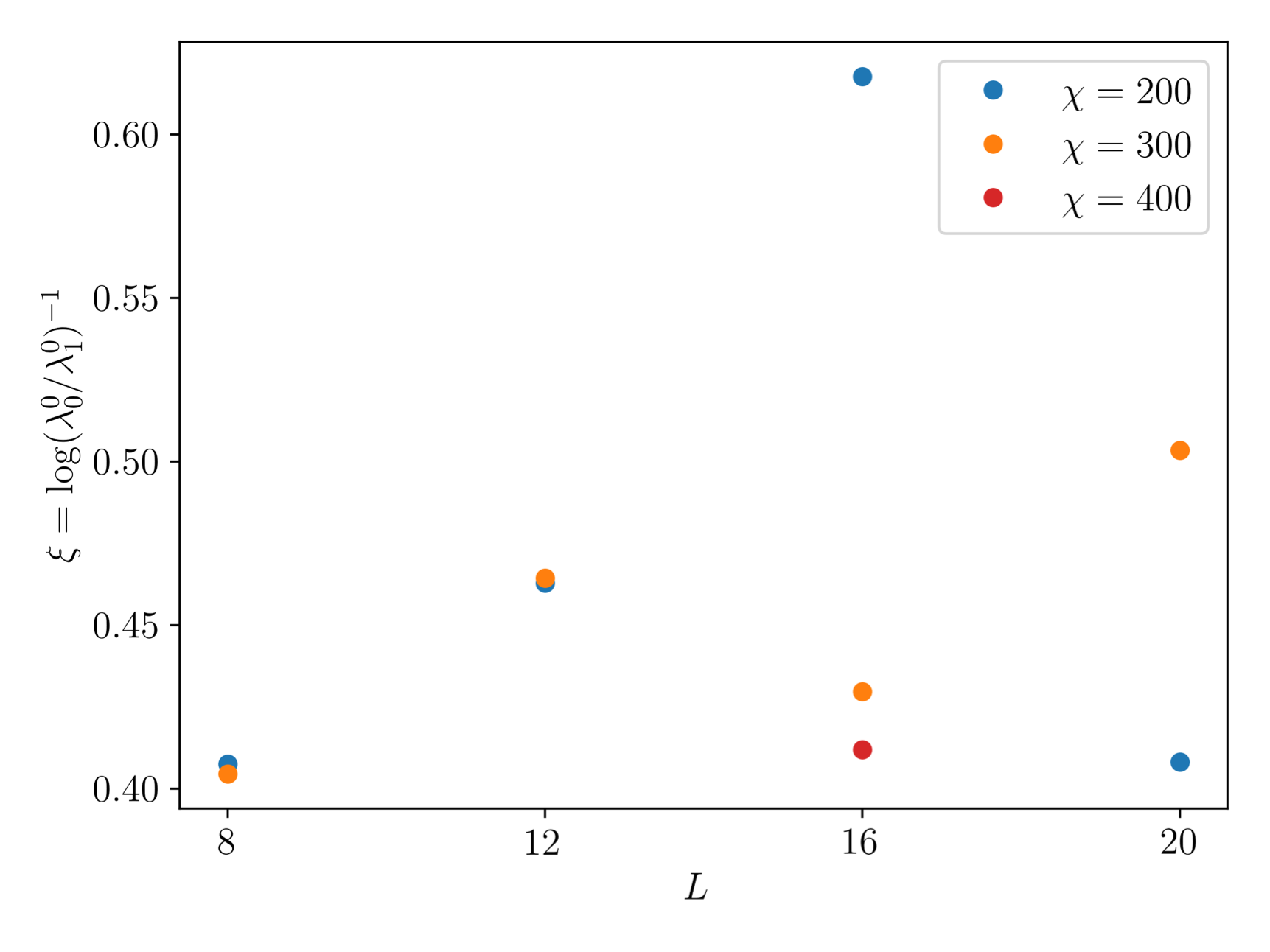}
    \caption{The correlation length, $\xi$, vs. cylinder circumference, $L$, for the equal weight tetramer superposition on the triangular lattice. For $L=8,12$ the bond dimensions $\chi = 200$ and $300$ are found to agree, while there is a large disagreement in correlation length for $L=16$. Going to larger bond dimension $\chi=400$ at $L=16$ appears to be closer to the $\xi$ value obtained at $\chi=300$, and appears consistent with a saturating correlation length. Unfortunately it was not feasible to run further analysis for $L=20$, where attempting $\chi = 400$ took several months before timing out.}
    \label{fig:corr_triang}
\end{figure}

Despite the numerical difficulty of finding the largest eigenvalues of ${\cal T}$ on the triangular lattice we were able to determine the upper bound on the correlation length for small cylinder circumferences and small bond dimensions. The results are shown in Fig.~\ref{fig:corr_triang}. While they appear consistent with a saturating correlation length as $L\rightarrow \infty$, and thus with a gapped $\mathbb{Z}_4$ TO, it is difficult to say anything precise given the disagreement between different bond dimensions at $L=16$. Further it is not feasible for us to attempt larger values of $L$ or $\chi$, as doing so would take several months to a year on our current hardware. As such we have left this problem for future work.

\end{document}